\documentclass[twocolumn,twoside,slac_two]{revtex4}
\usepackage{graphicx}
\usepackage{fancyhdr}
\usepackage{amsmath}
\newcommand\ensuretext[1]{\ensuremath{\text{#1}}}%
\newcommand\ergcms{\ensuretext{erg\,cm$^{-2}$\,s$^{-1}$}}%
\newcommand\phcms{\ensuretext{ph\,cm$^{-2}$\,s$^{-1}$}}%
\newcommand\phs{\ensuretext{ph\,s$^{-1}$}}%
\newcommand{\fermi}{\textit{Fermi}/LAT}
\newcommand\g{\ensuremath{\gamma}}%
\newcommand\rsn{\ensuremath{R_\mathrm{SN}}}%
\newcommand\mgas{\ensuremath{M_\mathrm{gas}}}%

\begin{document}

\title{Search for high-energy $\gamma$-ray emission from galaxies of the Local Group with \fermi\footnote{This manuscript derives from the work presented in \cite{2011arXiv1109.3321L}, in which more details can be found.}}

%

\author{J.-P. Lenain$^\ddagger$}
\affiliation{Landessternwarte, Universit{\"a}t Heidelberg, K{\"o}nigstuhl, 69117 Heidelberg, Germany ($^\ddagger$E-mail: jean-philippe.lenain@lsw.uni-heidelberg.de)}
\affiliation{ISDC Data Center for Astrophysics, Center for Astroparticle Physics, Observatory of Geneva, University of Geneva, Chemin d'Ecogia 16, CH-1290 Versoix, Switzerland}
\affiliation{LUTH, Observatoire de Paris, CNRS, Universit{\'e} Paris Diderot; 5 Place Jules Janssen, 92190 Meudon, France}
\author{R. Walter}
\affiliation{ISDC Data Center for Astrophysics, Center for Astroparticle Physics, Observatory of Geneva, University of Geneva, Chemin d'Ecogia 16, CH-1290 Versoix, Switzerland}

\begin{abstract}
Normal galaxies begin to arise from the shadows at high energies, as can be seen with the discovery of high-energy \g-ray emission from the Andromeda galaxy (M 31) by the \fermi\ collaboration.

We present a study on the search for high-energy emission around galaxies of the Local Group. No significant detection is found from the objects studied here. Upper limits on the high-energy emission of nearby normal galaxies are derived, and we discuss them in the context of \g-ray emission from cosmic ray interactions with the local interstellar medium in these galaxies.
\end{abstract}

\maketitle

\thispagestyle{fancy}


\section{INTRODUCTION}

Unusual suspects for extragalactic high-energy \g-ray emission begin to arise from the shadows: the \fermi\ collaboration recently reported the discovery of high-energy \g-ray from the Andromeda galaxy (M\,31), our neighbouring galaxy \cite{2010A+A...523L...2A}. More powerful sources, particularly active galactic nuclei (AGNs) with relativistic jets pointing close to the line of sight, the so-called blazars, represent the major part of extragalactic sources detected with \fermi\ (see e.g. \cite{2011arXiv1108.1420T}).

However, weaker sources such as starburst galaxies begin to be revealed as high-energy emitters, both in the GeV and TeV ranges, like M\,82 \cite{2010ApJ...709L.152A,2009Natur.462..770V} and NGC\,253 \cite{2010ApJ...709L.152A,2009Sci...326.1080A}. Starburst, Seyfert and normal galaxies have long been thought to marginally contribute to the extragalactic diffuse \g-ray emission (see e.g. \cite{1976MNRAS.175P..23S}), however recent studies tend to re-evaluate this contribution to be largely significant, if not dominant, compared to the emission from blazars (see e.g. \cite{2010ApJ...722L.199F}). We present here a study on the search for high-energy \g-ray emission from normal, nearby galaxies from the Local Group, in order to investigate any signature from cosmic rays interactions with the interstellar medium in the host galaxy, as opposed to the emission from AGNs.

\begin{figure*}
  \centering
  \includegraphics[width=0.8\textwidth]{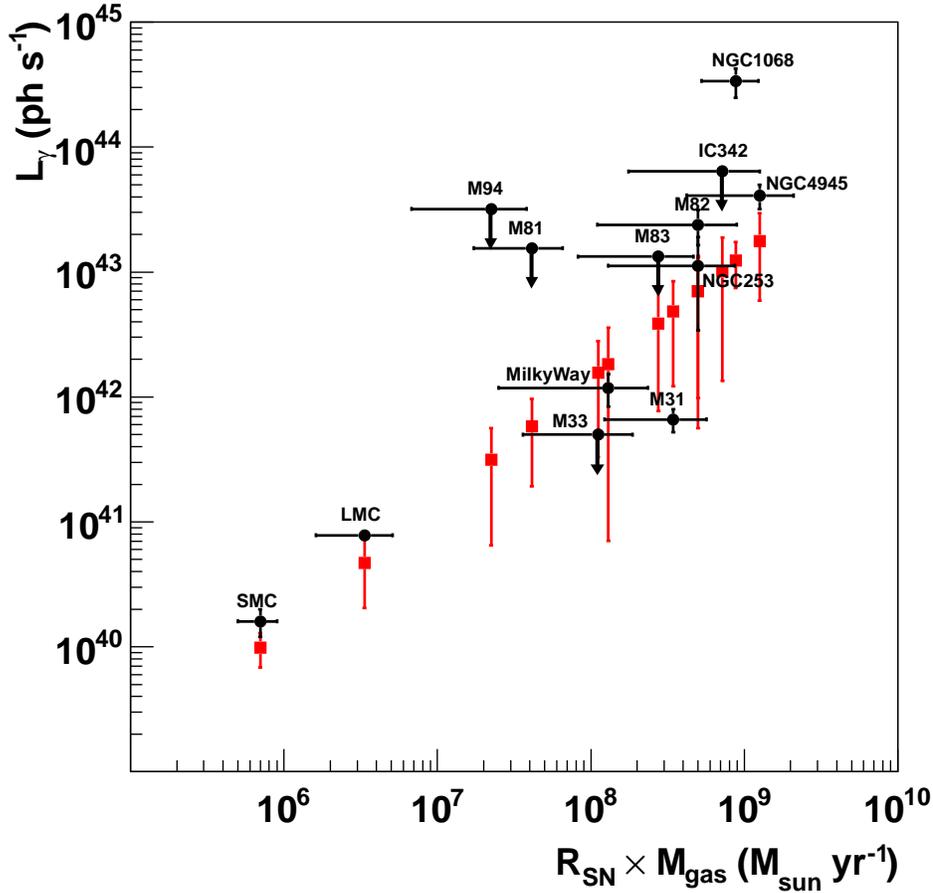}
  \caption{\g-ray luminosity, or the upper limit at 2$\sigma$ confidence level on the luminosity, of the different sources in the sample and other known high-energy emitting starburst and normal galaxies, given against the supernova rate times the total gas mass in these objects. The red square points show the expectations on the luminosity from the model of \cite{2001ApJ...558...63P}, accounting for the uncertainties on \rsn\ and \mgas.}
  \label{fig-relation}
\end{figure*}

\section{THE SAMPLE AND \textit{FERMI}/LAT DATA ANALYSIS}

High-energy emission from some galaxies of the Local Group has been reported for the small Magellanic cloud (SMC, \cite{2010A+A...523A..46A}), the large Magellanic cloud (LMC, \cite{2010A+A...512A...7A}), M\,31 \cite{2010A+A...523L...2A}, the starbursts M\,82 and NGC\,253 \cite{2010ApJ...709L.152A}, and Cen\,A \cite{2010Sci...328..725F,2010ApJ...719.1433A}. We study here the major galaxies from the Local Group for which no high-energy emission has been reported so far: M\,81, M\,83, IC\,342, Maffei\,1, Maffei\,2, and M\,94.

\begin{table*}
  \caption{Properties of galaxies from the Local Group.}
  \label{tab-ana_results}
  \centering
  \begin{tabular}{|l|c|c|c|c|}
    \hline
    Source         & d     & \rsn                         & \mgas           & $L_\g$                         \\
                   & kpc   & yr$^{-1}$                     & $10^9 M_{\odot}$   & $10^{43}$\,\phs                \\
    \hline
    Milky Way      & ...   & $0.02 \pm 0.01$              & $6.5 \pm 2.0$   & $0.12 \pm 0.03$               \\
    LMC            & 50    & $0.005 \pm 0.002$            & $0.67 \pm 0.08$ & $(7.8 \pm 0.8) \times 10^{-3}$ \\
    SMC            & 61    & $(1 \pm 0.2) \times 10^{-3}$  & $0.70 \pm 0.06$ & $(1.6 \pm 0.4) \times 10^{-3}$ \\
    M\,31          & 780   & $0.045 \pm 0.015$            & $7.66 \pm 2.38$ & $(6.6 \pm 1.4) \times 10^{-2}$ \\
    M\,33          & 847   & $0.050 \pm 0.015$            & $2.23 \pm 0.84$ & $< 5.0 \times 10^{-2}$         \\
    Maffei\,2      & 2800  & ?                            & ?               & $< 1.66$                      \\
    Maffei\,1      & 3010  & ?                            & ?               & $< 3.03$                      \\
    IC\,342        & 3280  & $0.18 \pm 0.10$              & $4.0 \pm 0.8$   & $< 6.40$                      \\
    M\,82          & 3530  & $0.2 \pm 0.1$                & $2.5 \pm 0.7$   & $2.39 \pm 0.75$               \\
    NGC\,4945      & 3600  & $0.3 \pm 0.2$                & $4.2$           & $4.09 \pm 0.92$               \\
    M\,81          & 3630  & $0.008 \pm 0.002$            & $5.16 \pm 1.72$ & $< 1.55$                      \\
    NGC\,253       & 3940  & $0.2 \pm 0.1$                & $2.5 \pm 0.6$   & $1.12 \pm 0.78$               \\
    M\,83          & 4500  & $0.050 \pm 0.025$            & $5.5 \pm 1.1$   & $< 1.33$                      \\
    M\,94          & 4660  & $0.04 \pm 0.02$              & $0.56 \pm 0.11$ & $< 3.19$                      \\
    NGC\,1068      & 14000 & $0.20 \pm 0.08$              & $4.4$           & $33.8 \pm 8.9$                \\
    \hline
  \end{tabular}
\end{table*}

A \fermi\ data analysis is performed for the sources in our sample using the public data from August 4, 2008 to January 1, 2011. We use the unbinned likelihood analysis \cite{2009ApJ...697.1071A} from the publicly available \textit{Science Tools} version \texttt{v9r18p6}. Data from the diffuse class events are selected, using the \texttt{P6\_V3} instrumental response functions, in the 200\,MeV--200\,GeV energy range. The isotropic \texttt{isotropic\_iem\_v02} model is used to account for both the extragalactic diffuse emission and residual instrumental background, while the model \texttt{gll\_iem\_v02} accounts for the contribution from the Galactic diffuse emission. The region of interest is taken from a circular region of 10$^\circ$ of radius around the nominal positions of the sources in our sample, while all the neighbouring sources up to 15$^\circ$ from the 11 months point source catalogue (1FGL, \cite{2010ApJS..188..405A}) are accounted for in the modelled reconstruction of the sources. The \texttt{gtlike} tool is used to assess the detection level of the studied objects, for all of which the Test Statistics (TS) are found to be below 25, resulting in no significant detection. Upper limits at 2$\sigma$ confidence level are thus derived on their flux in the 200\,MeV--200\,GeV energy range. No specific energy spectra are assumed to derive these limits, and the corresponding photon indices are left free to vary in the models. Table~\ref{tab-ana_results} presents a summary of our results for M\,81, M\,83, IC\,342, Maffei\,1, Maffei\,2, and M\,94, as well as previous results obtained on NGC\,1068 and NGC\,4945 \cite{2010A+A...524A..72L} and on the Milky Way, LMC, SMC, M\,31, M\,33, M\,82 and NGC\,253 (see \cite{2011arXiv1109.3321L} and references therein). More details on specific sources from the sample can be found in \cite{2011arXiv1109.3321L}.

\section{DISCUSSION}

Interactions between cosmic rays accelerated in stellar objects (e.g. supernov\ae remnants) and the ambient interstellar medium in starburst and normal galaxies are expected to generate high-energy \g-ray emission, through pion decay for hadronic cosmic rays as well as inverse Compton and bremsstrahlung radiations from leptonic cosmic rays (see e.g. \cite{2010ApJ...722L..58S}). \cite{2001ApJ...558...63P} suggested that the supernova rate \rsn\ and the total gas mass \mgas\ in a given galaxy could be a proxy for the expected \g-ray luminosity, through the relationship:

\begin{equation}
  \begin{aligned}
  F_\g (>100\,\mathrm{MeV}) = & 2.34 \times 10^{-8} \left(\frac{\rsn}{\rsn^\mathrm{MW}}\right) \left(\frac{\mgas}{10^8 M_{\odot}}\right)\\
   & \times \left(\frac{d}{100\,\mathrm{kpc}}\right)^{-2}\,\phcms
   \end{aligned}
\end{equation}
scaled with the supernova rate in the Milky Way \ensuremath{\rsn^\mathrm{MW}} and the distance of the considered galaxy $d$.

Figure~\ref{fig-relation} shows the \g-ray luminosity measured with \fermi\ for the known high-energy emitting starburst and normal galaxies, against $\rsn \times \mgas$, as well as the upper limits on the fluxes for the objects studied here which are not detected. The luminosities expected from the model of \cite{2001ApJ...558...63P} are also depicted in red, accounting for the observational uncertainties on \rsn\ and \mgas\ for the different objects.
For comparison, we also added in this plot the \g-ray luminosity for the Seyfert\,2 galaxy NGC\,1068 recently detected at high energies \cite{2010A+A...524A..72L}. NGC\,1068 also harbours a starburst region in its central part, but it was argued in \cite{2010A+A...524A..72L} that its high-energy emission more likely originates from the AGN activity.

It can be seen in Fig.~\ref{fig-relation} that the upper limit reported on M\,33 by the \fermi\ collaboration \cite{2010A+A...523L...2A} is very stringent compared to expectation, if the high-energy flux is truly directly connected to $\rsn \times \mgas$, i.e. if the high-energy emission is only due to cosmic ray interactions with the interstellar medium, without any other contribution. In this case, a detection of M\,33 with \fermi\ should be imminent. A non-detection of M\,33 in the future years could be an indication for e.g. a particularly efficient electron escape in this galaxy, partly suppressing the high-energy emission. It can be noted from Fig.~\ref{fig-relation} that the same applies to M\,83, which should also be detectable soon with \fermi. Apart from the peculiar case of NGC\,1068 for which the \g-ray emission may be dominated by the AGN activity, the observations for the other sources reported here are fully consistent with expectations from the model of \cite{2001ApJ...558...63P}.

An alternative scenario can be invoked for expected high-energy emission from normal galaxies. Indeed \cite{2010ApJ...717..825D} recently reported the discovery of giant \g-ray bubbles in the inner Milky Way, and an extensive discussion on different possible mechanisms for their origin can be found e.g. in \cite{2010ApJ...724.1044S}. The emission from these giant bubbles more likely comes from inverse Compton scattering of a hard population of Galactic cosmic ray electrons, the latter may also be at the origin of the microwave haze seen with \textit{WMAP} \cite{2008ApJ...680.1222D}. If such emission from giant bubbles occurs in other normal galaxies, one could expect to detect that signal in galaxies from the Local Group, in addition to the emission discussed above, which would result in a higher \g-ray flux compared to what is expected from the model of \cite{2001ApJ...558...63P}. Such hypothesis can already be ruled out in the peculiar case of M\,33, owing to the constraining upper limit reported on its \g-ray flux.

High-energy emission could also arise from a third hypothesis. Collective emission from galactic sources within the host galaxies, such as pulsars, pulsar wind nebul\ae\ or supernova remnants, could be detected in nearby galaxies. To evaluate the contribution of such emission, we estimate the expected flux for this collective emission using the \g-ray luminosities for pulsars and pulsar wind nebul\ae\ currently detected with \fermi\ (see \cite{2011arXiv1109.3321L} for more details), assuming these galactic objects are located in a hypothetical galaxy distant by 3\,Mpc, which is the mean distance for the sources in our sample. The modelled flux was intentionally over-estimated by assuming 1000 such sources in the host galaxy, which is much more numerous than the $\sim$60 similar objects currently detected with \fermi\ in our Galaxy \cite{2010ApJS..187..460A,2011ApJ...726...35A}. Even with such optimistic conditions, the collective flux from pulsars is computed to be $F(E>100\,\mathrm{MeV}) \sim 5 \times 10^{-14}$\,\ergcms, and the one for pulsar wind nebul\ae\ amounts to $F(E>100\,\mathrm{MeV}) \sim 10^{-13}$\,\ergcms. These estimates are well below the sensitivity achievable by \fermi\ within 2 years of observations\,\footnote{see \url{http://www.slac.stanford.edu/exp/glast/groups/canda/lat_Performance.htm}.}. The contribution from collective emission of galactic objects in the host galaxy is thus found to be negligible compared to the one from cosmic ray interactions with the local interstellar medium.

\section{CONCLUSIONS}

High-energy \g-ray emission from normal galaxies of the Local Group has been searched for. No normal galaxy has been detected at high energy yet, apart from M\,31 \cite{2010A+A...523L...2A}. We derived upper limits on their flux, which are found to be fully compatible with expectations from the model of \cite{2001ApJ...558...63P} for cosmic ray interactions with the local interstellar medium. If this model holds, a detection should be imminent with \fermi\ for M\,33 and M\,83. The expected \g-ray flux of the other sources in the sample are below the sensitivity of \fermi, in accordance with \cite{2010ApJ...722L.199F} who estimated that $\sim$5 normal galaxies should be individually detectable in the \fermi\ lifetime.

Deeper observations with \fermi\ will possibly lead to the detection of a few other normal, non-active galaxies in the near future.

\bigskip 
\begin{acknowledgments}
This research has made use of NASA's Astrophysics Data System (ADS), of the SIMBAD database, operated at CDS, Strasbourg, France, and of the NASA/IPAC Extragalactic Database (NED) which is operated by the Jet Propulsion Laboratory, California Institute of Technology, under contract with the National Aeronautics and Space Administration.

J.-P. L. would like to warmly thank Dr. Christian Farnier and Dr. Andrea Tramacere for invaluable help and discussions during the development of this work.
\end{acknowledgments}

\bigskip 

\end{document}